\documentclass[final,5p,times,twocolumn,square,numbers,comma]{elsarticle}
\usepackage{silence}
\usepackage{amssymb}
\usepackage{amsmath}
\usepackage{lipsum}
\usepackage{booktabs}
\usepackage{multirow}
\usepackage{amsfonts}
\usepackage{graphicx}
\usepackage{xcolor}
\usepackage{transparent}
\usepackage{ragged2e}
\usepackage{amsthm}
\usepackage[hidelinks]{hyperref}
\usepackage{tikz}
\usepackage{booktabs}
\usepackage{array} 
\usepackage{longtable}
\usepackage[export]{adjustbox} 
\usepackage{algorithm}
\usepackage{algpseudocode}
\usepackage{tabularx}
\usepackage{float} 
\usepackage{afterpage} 
\usepackage{enumitem}

\makeatletter
\def\ps@pprintTitle{%
    \let\@oddhead\@empty        
    \let\@evenhead\@empty       
    \def\@oddfoot{\reset@font\hfil\@date}
    \let\@evenfoot\@oddfoot     
}
\makeatother

\begin{document}
\begin{frontmatter}

\title{Beyond Wave Variables: A Data-Driven Ensemble Approach \\ for Enhanced Teleoperation Transparency and Stability}

\author[usthb,oulu]{Mitiche Nour}
\author[usthb]{Ferguene Farid}
\author[oulu]{Oussalah Mourad}

\author{ \\ \textit{nmitiche@usthb.dz, fferguene@usthb.dz, mourad.oussalah@oulu.fi}}

\affiliation[usthb]{organization={LRPE Laboratory, Automatic Department},
                addressline={U.S.T.H.B University BP n°32, El Alia-Bab Ezzouar}, 
                postcode={16111 Algiers}, 
                country={Algeria}}

\affiliation[oulu]{organization={Center for Machine Vision and Signal Processing, University of Oulu},
                addressline={Pentti Kaiteran katu 1},
                city={Oulu},
                postcode={90570},
                country={Finland}}

\begin{abstract}
Time delays in communication channels present significant challenges for bilateral teleoperation systems, affecting both transparency and stability. Although traditional wave variable-based methods for a four‑channel architecture ensure stability via passivity, they remain vulnerable to wave reflections and disturbances like variable delays and environmental noise. This article presents a data‑driven hybrid framework that replaces the conventional wave‑variable transform with an ensemble of three advanced sequence models, each optimized separately via the state‑of‑the‑art Optuna optimizer, and combined through a stacking meta‑learner. The base predictors include an LSTM augmented with Prophet for trend correction, an LSTM–based feature extractor paired with clustering and a random forest for improved regression, and a convolutional‑LSTM model for localized and long‑term dynamics. Experimental validation was performed in Python using data generated from the baseline system implemented in MATLAB/Simulink. The results show that our optimized ensemble achieves a transparency comparable to the baseline wave-variable system under varying delays and noise, while ensuring stability through passivity constraints.
\end{abstract}


\end{frontmatter}

\section{Introduction}
\label{introduction}
Bilateral teleoperation systems allow human operators to intuitively handle remote robotic manipulators, while receiving haptic feedback. This kind of system is widely used in critical tasks such as space exploration\cite{space}, deep-sea and hazardous environment operations \cite{Sheridan1992, udw}, as well as more delicate applications, such as telesurgery \cite{endo,millirob,wan}. One of the main challenges related to these systems is the presence of time delays within communication channels, which may compromise the system's transparency, and therefore degrade both position tracking accuracy and reliability of the force feedback, resulting in system instability \cite{Lawrence1993}. 

The four-channel control architecture \cite{Lawrence1993}, significantly enhanced by Chen \textit{et al.} \cite{Chen2018}, has conventionally addressed the time delay issues by integrating a modified wave-variable transform \cite{Niemeyer1991}. This approach ensures stability through passivity constraints and enhances transparency by mitigating wave reflections within the communication channel. However, the limitation of this methodology is that it assumes constant delays and mostly linear system dynamics, restricting its robustness.  It is particularly vulnerable when confronted with real-world scenarios where it must handle the complex, non-linear behaviors introduced by variable communication delays, stochastic signal noise, or unexpected environmental changes.

The recent advances in machine learning, particularly those related to Long Short-Term Memory (LSTM) neural networks \cite{Hochreiter1997}, offer promising insights to model complex non-linear dynamics found in teleoperation systems. However, real-world communication channels are affected by multiple challenges, including variable delays, stochastic noise, and nonlinearities. A single model architecture may struggle to address all these challenges comprehensively. A model adept at long-term trend prediction might fail to capture localized noise, and vice-versa. For instance, a pure LSTM might excel at modeling smooth operator motions but struggle with high-frequency jitter from network noise, whereas a convolutional architecture may capture local disturbances but miss the broader temporal context. Intuitively, this trade-off in model specialization motivates an ensemble approach, where the diverse strengths of multiple models can be leveraged for superior overall performance and robustness.

For this purpose, we propose a data-driven communication channel that replaces the conventional wave-variable transform by a robust learning-based surrogate. Our framework is based on a stacking ensemble strategy \cite{Wolpert1992}, which combines the strengths of multiple hybrid deep learning architectures and specialized base models, each optimized independently. These models' predictions are then intelligently integrated by a higher-level meta-learner. A critical aspect of integrating any learned model into a control loop is the guarantee of stability. Therefore, our proposed model is not only optimized for predictive accuracy, but also explicitly verified to maintain the overall system passivity, ensuring that it would not introduce energy into the system, and thereby preserving stability.

\begin{figure*}[h]
    \centering
    \includegraphics[width=0.85\linewidth]{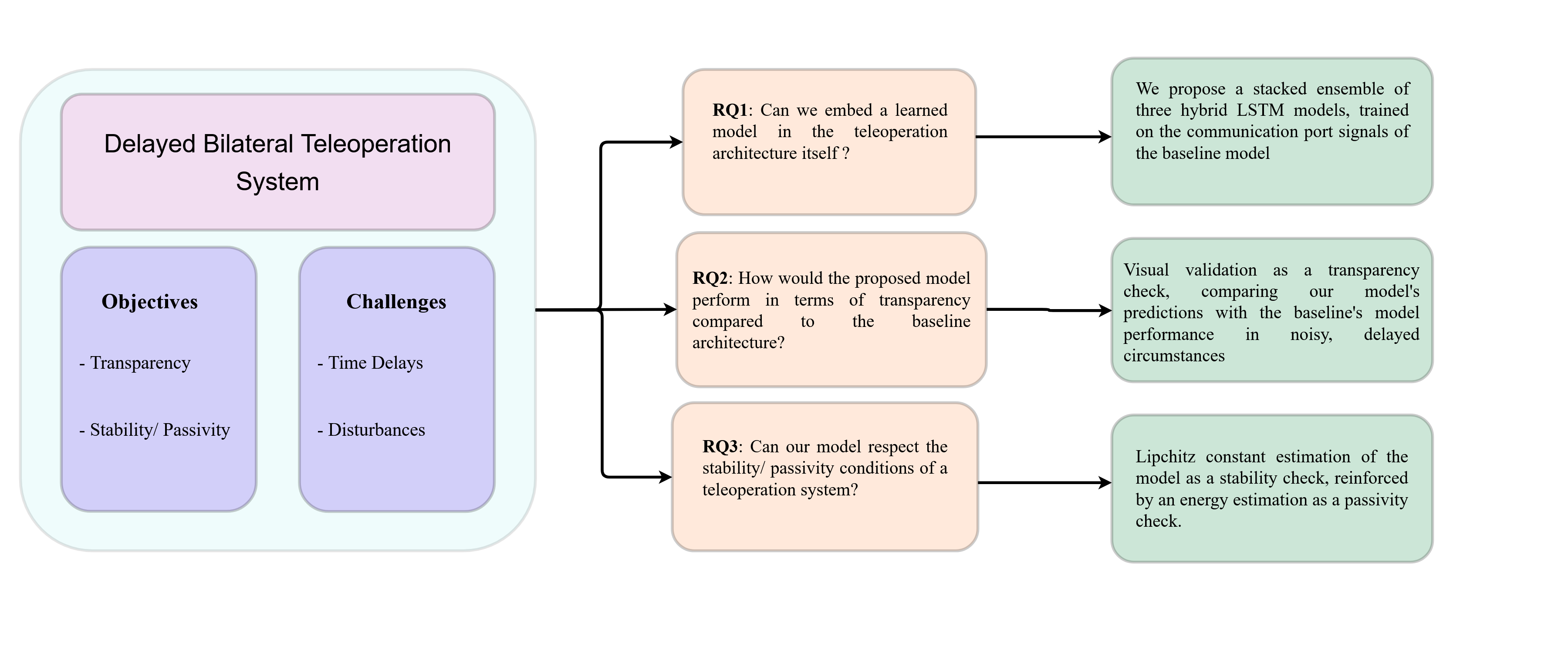}
    \caption{Overview of our problematic, research questions and the proposed methodology.}
    \label{overall}
\end{figure*}

To guide our investigation, as illustrated in \textbf{\autoref{overall}}, this study addresses the following key research questions:

\begin{itemize}
    \item \textbf{\textit{RQ1:}} Can a data-driven ensemble of hybrid deep learning models serve as a viable and robust replacement for the traditional wave-variable transform in a four-channel teleoperation architecture?
    
    \item \textbf{\textit{RQ2:}} How does the performance of such an ensemble, in terms of transparency and robustness to variable time delays and noise, compared to the established wave-variable baseline?

    \item \textbf{\textit{RQ3:}} Can this complex, learned model be formally verified to uphold the strict passivity and stability requirements necessary for safe and reliable human-robot interaction in a teleoperation loop?

    \item[] 
\end{itemize}

The primary contributions of this work are threefold. First, we propose a novel data-driven communication channel for a four-channel teleoperation architecture, replacing the traditional wave-variable transform with a stacking ensemble of three distinct, optimized hybrid deep learning models. Secondly, we introduce a rigorous validation framework that not only evaluates predictive accuracy, but also formally analyzes the stability of the learned model through Lipschitz constant estimation and passivity verification, ensuring its suitability for teleoperation control architectures. Finally, through extensive simulation, we demonstrate that our ensemble model achieves transparency comparable to the baseline under ideal conditions while offering superior robustness to variable time delays and signal noise, a common challenge in real-world applications.

The remainder of this paper is organized as follows. Section 2 provides a comprehensive review of the four-channel baseline architecture and discusses related work on delayed teleoperation systems. Section 3 details the proposed hybrid LSTM-based methodology and Optuna-driven optimization. Section 4 presents the experimental setup, simulation results, and comparative analysis. Section 5 discusses the implications, limitations, and future directions. Finally, Section 6 concludes the paper by summarizing the key contributions.

\section{Background and Related Work}
The main goal of bilateral teleoperation systems is to extend human presence and manipulation abilities to remote or hazardous environments \cite{Hokayem2006}. This can be achieved by creating a closed-loop human-robot interface that allows an operator to control a distant robot (slave), and receive in return a haptic feedback from the remote environment \cite{Sheridan1992}.

While offering immense potential, the challenges of remote operation, particularly communication delays, have led decades of research towards robust and transparent control strategies \cite{Lawrence1993}. In this section, we will start by explaining the foundational four-channel control architecture \cite{Lawrence1993}, which serves as the baseline for this study, outlining its principles, strengths and limitations. Then, we will review some contributions in time delay compensation in teleoperation systems, notably traditional techniques such as wave-variable control \cite{Niemeyer1991} and passivity-based approaches \cite{Anderson1988}, with a particular focus on the role of data-driven and machine learning approaches \cite{Sun2020}.

\subsection{Wave-Variable based Bilateral Teleoperation}
The four-channel control architecture \cite{Lawrence1993} has become a widely adopted framework for bilateral teleoperation for its ability to decouple force and velocity/position feedback, offering a more intuitive and stable control experience \cite{Hokayem2006}. This architecture contains four distinct feedback loops: two position/velocity feedback loops (master-to-slave and slave-to-master) and two force feedback loops (master-to-slave and slave-to-master). This decoupling allows us to adjust both the position/velocity tracking and the force reflection, which are essential to achieve a good transparency.

\begin{figure}[t]
    \centering
    \includegraphics[width=\linewidth]{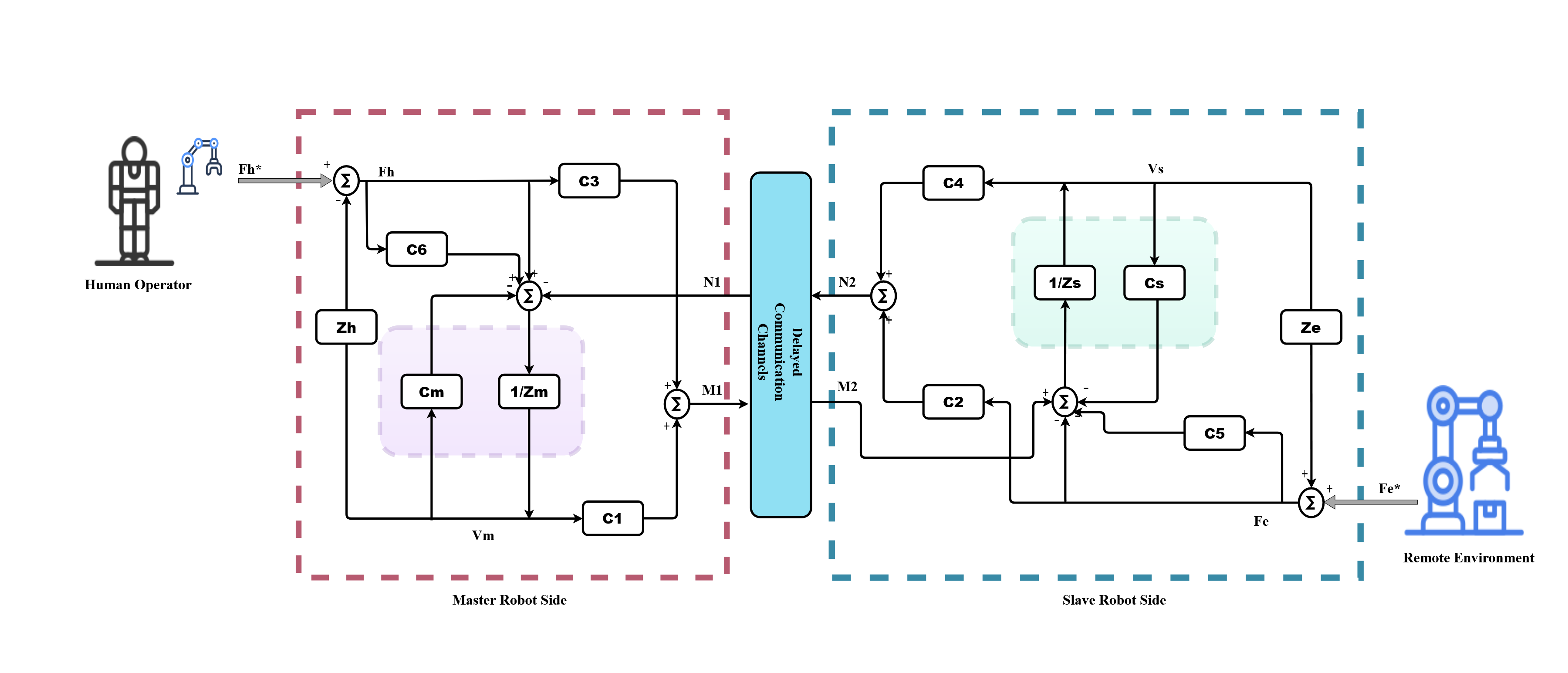} 
    \caption{The Improved four-channel teleoperation architecture, adapted from Chen \textit{et al.} \cite{Chen2018}}
    \label{sys}
\end{figure}

\begin{figure}[H]
    \centering
    \includegraphics[width=\linewidth]{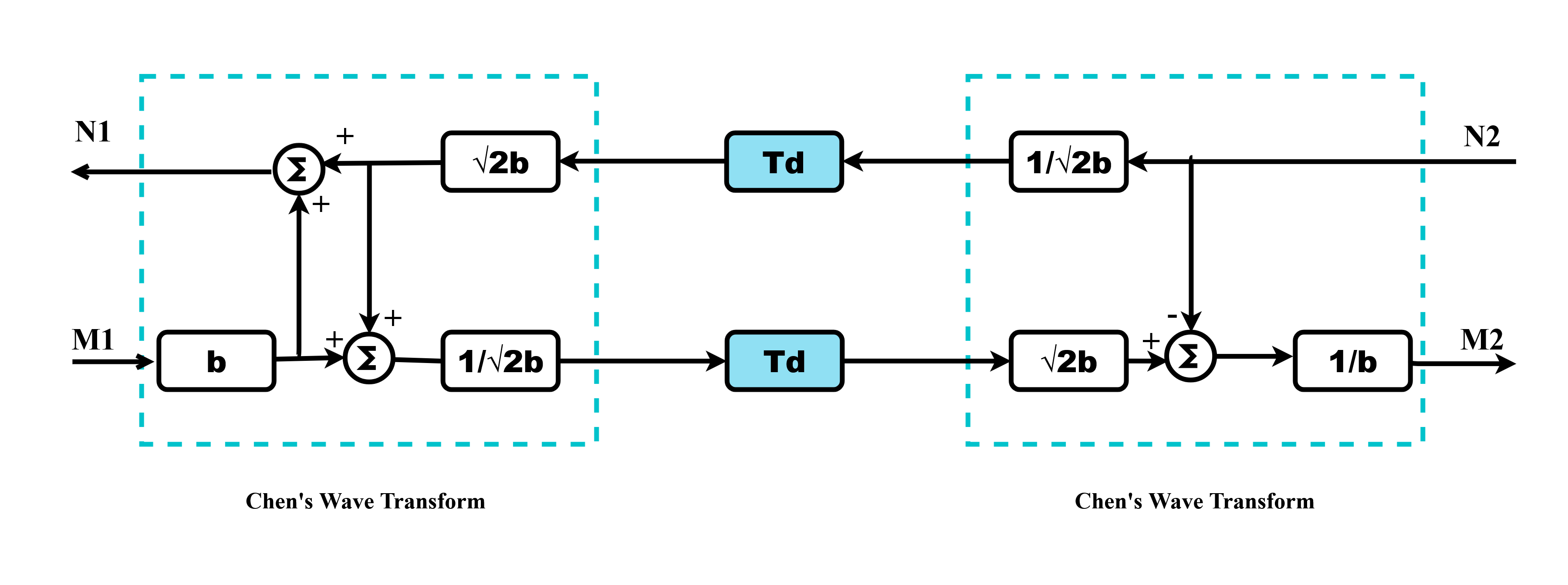}
    \caption{Compensated Communication Channel with Modified Wave Transform.\cite{Chen2018}}
    \label{cha}
\end{figure}

The key principle to achieve stability in delayed teleoperation within a four-channel framework is the concept of passivity \cite{Anderson1988}. A passive system cannot generate energy, it can only store or dissipate it. However, time delays can induce non-passive behavior in communication channels, leading to instability, oscillations, and a degraded sense of presence for the operator \cite{Niemeyer1991}. To address this, Chen \textit{et al.} \cite{Chen2018} enhanced the four-channel architecture, as shown in \textbf{\autoref{sys}}, by integrating a modified wave variable transform, detailed in \textbf{\autoref{cha}}. The wave variable transform is a widely recognized technique that converts conventional force and velocity signals into \textit{wave variables} that propagate along the communication channel \cite{Niemeyer1991}. 

This transformation, based on the scattering theory, creates a passively-behaved communication channel. Chen et al.'s enhancement involved adjusting the scaling parameter $'b'$ of the wave transform and redesigning the local force and position controllers ($C_1$ through $C_6$) to optimize the trade-off between stability and transparency for a given delay. An optimization framework has been proposed later on by Mitiche \textit{et al.} \cite{mitiche23} using Simulink's dedicated Optimization Toolbox.

\nocite{Nuno2011,Smith1957,NormeyRico2008, Ryu2004,Greff2017,lstm2,Liu2023,Lee20,Wang21,Wang2024}

By ensuring that the energy flowing into the communication channel is always greater than or equal to the energy flowing out, the wave-variable method guarantees the stability of the teleoperation system, even in the presence of constant communication delays \cite{Niemeyer1991}.

\begin{table*}[t!]
\centering
\footnotesize 
\setlength{\tabcolsep}{4pt} 
\renewcommand{\arraystretch}{0.75}
\caption{Analysis of Standalone Time-Series Techniques in the context of Delayed Teleoperation.}
\label{tab:standalone_analysis_ultra_compact}
\begin{tabular*}{\textwidth}{@{\extracolsep{\fill}}
    >{\centering\arraybackslash}p{3.5cm} 
    >{\raggedright\arraybackslash}p{6.5cm} 
    >{\raggedright\arraybackslash}p{6.5cm} 
@{} }
\toprule
\textbf{Technique} & \textbf{Mechanism in Teleoperation Time-Series} & \textbf{Limitations} \\
\midrule

\multirow{4}{*}{\centering LSTM Network \cite{lstm1, lstm2}} &
\begin{itemize}[nosep] 
    \item Models temporal dependencies via an internal memory cell.
    \item Learns dynamic master-slave relationships over time.
    \item Predicts future slave states to compensate for delay.
\end{itemize} &
\begin{itemize}[nosep]
    \item Computationally expensive for real-time use.
    \item Prone to overfitting on small or non-diverse datasets.
    \item Poor generalization to unseen environmental dynamics.
    \item "Black-box" nature complicates safety verification.
\end{itemize} \\
\midrule

\multirow{3}{*}{\centering Prophet \cite{chen2023analysis}} &
\begin{itemize}[nosep]
    \item Decomposes signals into trend and seasonality components.
    \item Predicts periodic patterns in highly repetitive tasks.
\end{itemize} &
\begin{itemize}[nosep]
    \item Unsuitable for non-stationary and aperiodic signals.
    \item Additive model fails to capture non-linear interactions.
    \item Cannot react to rapid changes; unfit for delay compensation.
\end{itemize} \\
\midrule

\multirow{4}{*}{\centering K-means Clustering \cite{km,km2,km3}} &
\begin{itemize}[nosep]
    \item Segments data into distinct operational phases (e.g., contact).
    \item Enables use of phase-specific controllers or predictors.
\end{itemize} &
\begin{itemize}[nosep]
    \item Not a predictive model; only categorizes data points.
    \item Ignores the temporal sequence of the data.
    \item Requires predefining the number of clusters ($k$).
    \item Sensitive to initialization and non-spherical clusters.
\end{itemize} \\
\midrule

\multirow{3}{*}{\centering Random Forest Regression \cite{rf}} &
\begin{itemize}[nosep]
    \item Regresses the next value from a window of past data.
    \item Captures non-linear relationships without parametric assumptions.
\end{itemize} &
\begin{itemize}[nosep]
    \item Cannot extrapolate beyond the training data range.
    \item Lacks a true temporal state; predictions are independent.
    \item Memory and computationally intensive with large forests.
\end{itemize} \\
\midrule

\multirow{3}{*}{\centering CNN (1D) \cite{cnn1,cnl4}} &
\begin{itemize}[nosep]
    \item Applies 1D convolutions to learn local patterns/features.
    \item Detects key motifs (e.g., initial contact) for prediction.
\end{itemize} &
\begin{itemize}[nosep]
    \item Requires a fixed-size input window, limiting flexibility.
    \item Weaker at modeling long-range dependencies than LSTMs.
    \item Performance is sensitive to the input's time scale.
\end{itemize} \\
\bottomrule
\end{tabular*}
\end{table*}

This passivity constraint also improves transparency by actively absorbing or reflecting undesirable energy, thereby reducing phenomena like wave reflection and enhancing the operator's perception of the remote environment.

Despite its significant contribution to achieve stability and improve transparency, the traditional wave variable-based four-channel architecture, as developed by Chen \textit{et al.} \cite{Chen2018}, operates under some simplifying assumptions: it considers the time delays constant and often presumes linear system dynamics \cite{Nuno2011}. These assumptions impose considerable limitations in real-world applications, as time delays are generally variable (due to network congestion, fluctuating distances, or varying computational loads), and system dynamics are often non-linear \cite{Nuno2011}. Thus, this traditional approach often demonstrates reduced robustness when confronted with when confronted with the non-linearities introduced by unpredictable signal noise or dynamic delay fluctuations, resulting in degraded performance and, in some cases, instability. This susceptibility to disturbances highlights the necessity for more adaptive and robust delay compensation strategies.

\subsection{Related Work}
The challenge of time delays in teleoperation has spurred a variety of research techniques beyond wave variables, each with their own advantages and limitations. Early approaches included the Smith Predictor, which attempts to compensate for known delays by predicting future system states \cite{Smith1957}, but struggles significantly with variable delays and model inaccuracies \cite{NormeyRico2008}. Robust control techniques have been applied to design controllers that are resilient to uncertainties, such as bounded delays based methods \cite{Ryu2004}, but often at the expense of transparency or higher system modeling complexity. Other strategies involve observer-based techniques which estimate remote states, and various forms of predictive control, which anticipate system behavior to mitigate delay effects \cite{Nuno2011}. While these methods offer valuable insights, many still face difficulties in efficiently handling moderate to high stochastic delays, or they require precise knowledge of the system's dynamic model, which can be difficult to obtain and maintain in practice.

More recently, the widespread success of deep learning techniques with handling complex, non-linear and sequential data has opened new avenues for teleoperation research \cite{Sun2020}. Recurrent Neural Networks (RNN) based techniques, more specifically Long Short-Term Memory (LSTM) networks \cite{Hochreiter1997}, are exceptionally well-suited for modeling time-series data and capturing long-term dependencies, making them promising candidates for predicting delayed signals or modeling complex communication dynamics \cite{Greff2017}. LSTMs have demonstrated strong performance across diverse control and prediction tasks, including system identification, fault detection, and predictive maintenance. In the context of teleoperation, novel approaches have investigated  LSTMs for trajectory and force prediction, as well as adaptive control using reinforcement learning or neural network-based compensation schemes \cite{lstm2}. For instance, LSTMs have been explored to design predictive controllers that anticipate master or slave dynamics, enhance robustness to network-induced delays, and even reduce physiological tremors through signal filtering \cite{Liu2023,Lee20}. Moreover, LSTM-based adaptive controllers have been proposed to cope with unmodeled dynamics, such as friction or disturbance forces, and to enable event-triggered updates in bandwidth-constrained scenarios \cite{Wang21,Wang2024}. 

However, as we previously mentioned, despite the considerable power of standalone LSTM networks in handling specific aspects of teleoperation, the inherent complexity and multi-faceted nature of real-world communication channels often present challenges that a single architectural paradigm cannot fully address, which motivates a move towards more sophisticated architectures that can leverage the unique strengths of different models in a complementary fashion.

Consequently, some initiatives have turned to hybrid and ensemble-based learning methods that could incorporate the strengths of different predictive models. Combining techniques such as LSTM \cite{LSTM}, Prophet \cite{Taylor2018}, K-means clustering \cite{KMeans}, Random Forest regression \cite{Breiman2001}, and CNN \cite{cnn0, Zhao2017, CNN}, may offer better modeling flexibility and robustness. The \textbf{\autoref{tab:standalone_analysis_ultra_compact} }details each of these models' approach in a teleoperation system, as well as their limitations as a standalone technique.

Furthermore, the use of state-of-the-art hyperparameter optimization tools such as Optuna \cite{Akiba2019} ensures that these architectures are efficiently and consistently fine-tuned for optimal performance. Stacked generalization (stacking) \cite{Wolpert1992} offers a compelling meta-learning strategy to aggregate the strengths of multiple base models (level-0 learners), and has proven useful in reducing variance and improving generalization in dynamic environments. However, the integration of such ensemble techniques directly within a passivity-constrained teleoperation framework remains largely unexplored.

This study addresses that gap by proposing an ensemble of three optimized hybrid LSTM models, integrated into the communication channel of the four-channel architecture, maintaining passivity while improving robustness to disturbances and varying delays. Each architecture is optimized using Optuna \cite{Akiba2019}, a state-of-the-art hyperparameter optimization framework that enables consistent tuning across model variants. To fully exploit their complementary strengths, the three optimized hybrid models are employed as base learners in a stacked ensemble, where a meta-learner synthesizes their outputs to generate the final delay-compensated signal \cite{Wolpert1992}. This stacked architecture serves as a direct replacement for the wave-variable transform, operating within the four-channel bilateral teleoperation framework.

\begin{figure*}[ht!]
    \centering
    \includegraphics[width=0.85\textwidth]{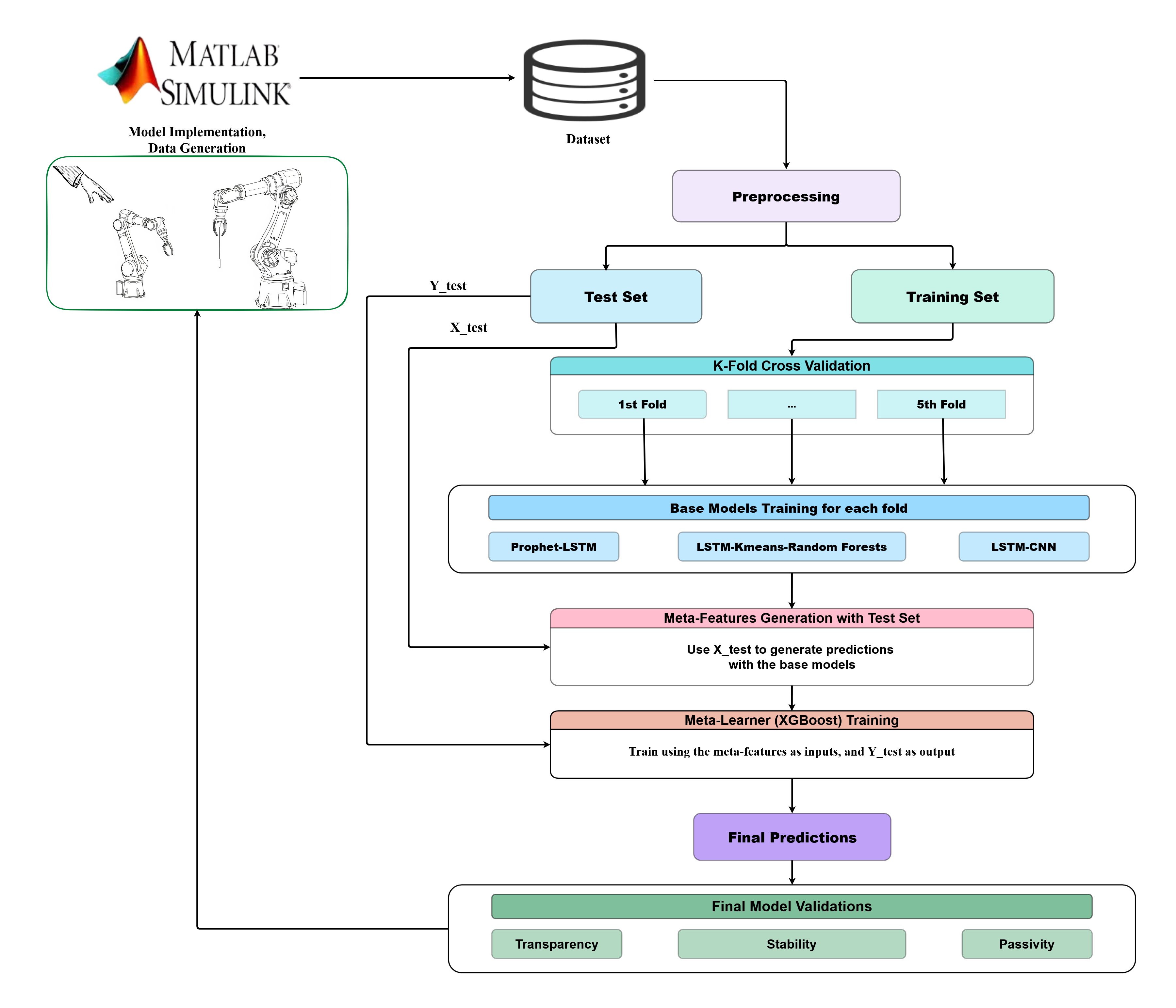}
    \caption{The complete data processing and model training pipeline. The workflow begins with data generation in MATLAB/Simulink, followed by preprocessing and splitting. Base models are trained using K-fold cross-validation to generate meta-features, which are then used to train the final meta-learner.}
    \label{fig:pipeline}
\end{figure*}

While learned models have been widely used to enhance local predictions or adaptive control, few studies have attempted to embed one directly into the inter-robot communication pathway while preserving formal stability guarantees. By explicitly constraining the system to meet passivity requirements, our approach maintains the stability and robustness of classical architectures while extending their performance in the face of dynamic and non-deterministic delays.

\section{Methodology}
This section details the proposed data-driven framework illustrated in \textbf{\autoref{fig:pipeline}}. We first describe the baseline system and data generation process. Then, we present the architecture of the stacking ensemble, and detail its training procedure. Finally, we outline the metrics and procedures used for performance and stability evaluation.

\subsection{Baseline System Modeling and Dataset Generation}
The foundation of this study is a simulation of a four-channel bilateral teleoperation system. This model, based on the wave-variable control architecture of Chen \textit{et al.} \cite{Chen2018}, serves as the baseline framework. It captures the master-slave dynamics, communication delays, and force feedback mechanisms while ensuring passivity under ideal noise-free conditions. 

To generate a robust training and validation dataset, the overall system was subject to a diverse set of input signals, $F^*_h$, including steps, sinusoids, and splines-generated signals, in both ideal and disturbed communication scenarios. The targeted ideal signals were stored as ground truth for our supervised learning.

\subsection{Hybrid LSTM Integration and Robustness Paradigm}
The core contribution of this work lies in the use of a meta model directly within the communication channel of a teleoperation framework, as a learned alternative to traditional wave-variable transforms. Rather than relying on a single model, we propose a hybrid stacking ensemble composed of three optimized LSTM-based architectures:

\begin{itemize}
        \item \textbf{Prophet-LSTM}: When it comes to delayed data, Prophet, as a decomposable time-series model designed to handle data with strong seasonal effects and trend changes \cite{Taylor2018}, is particularly good at capturing the macro-dynamics while preserving robustness to missing data points and trends shifts. On the other hand, LSTM models are specifically designed to learn long-term dependencies in sequential data \cite{LSTM}. 
        
        Inspired by works in other research domains, e.g., stock market forecasting \cite{prlstm1, prlstm4},  electricity load forecasting \cite{prlstm2} and aviation risk prediction \cite{prlstm3}, the suggested  Prophet-LSTM leverages a "divide and conquer" strategy that combines the temporal decomposition capabilities of Prophet with LSTM to correct systematic residuals in delay prediction. The architecture of the proposed Prophet-LSTM is detailed in \textbf{Algorithm \ref{alg:prophet_lstm}}.

    \begin{algorithm}[t] 
    \caption{Prophet-LSTM Architecture}
    \label{alg:prophet_lstm}
    \begin{algorithmic}[1]
    \Procedure{Fit}{$X_{train}, y_{train}$}
        \For{each output dimension $d$}
            \State Train a Prophet model $M_{prophet, d}$ on $(X_{train}, y_{train,d})$ using $X_{train}$ as exogenous regressors.
            \State Calculate predictions $P_{prophet, d} \gets M_{prophet, d}.\text{predict}(X_{train})$.
            \State Calculate the residuals: $R_{train, d} \gets y_{train,d} - P_{prophet, d}$.
            \State Train an LSTM model $M_{lstm, d}$ to predict $R_{train, d}$ from $X_{train}$.
        \EndFor
    \EndProcedure
    \Procedure{Predict}{$X_{test}$}
        \State For each output dimension $d$:
        \State \quad Generate base predictions from the trained Prophet model: $P_{prophet, d}$.
        \State \quad Generate residual predictions from the trained LSTM model: $P_{residuals, d}$.
        \State Combine predictions for all dimensions: $P_{final} \gets P_{prophet} + P_{residuals}$.
        \State \textbf{return} $P_{final}$.
    \EndProcedure
    \end{algorithmic}
    \end{algorithm}
    
    \item[]

    \item \textbf{CNN-LSTM}: Inspired by existing hybridization schemes for time-series forecasting tasks across many domains \cite{cnl1, cnl2, cnl3,cnl4}, we suggest a simple CNN-LSTM architecture that uses 1D convolutional layers for local pattern detection, followed by LSTM for long-term temporal dependencies \cite{Zhao2017,hand}. The architecture is detailed in \textbf{Algorithm \ref{alg:cnn_lstm}}.

    \item[]

    \begin{algorithm}[H]
    \caption{CNN-LSTM Architecture}
    \label{alg:cnn_lstm}
    \begin{algorithmic}[1]
    \Procedure{Fit}{$X_{train}, y_{train}$}
        \State Train a unified CNN-LSTM model on $(X_{train}, y_{train})$.
        \State \textit{The architecture consists of Conv1D and MaxPooling layers for feature extraction followed by an LSTM layer for sequence processing, and finally Dense layers for output.}
    \EndProcedure
    \Procedure{Predict}{$X_{test}$}
        \State Generate predictions using the single, trained CNN-LSTM model.
        \State \textbf{return} predictions.
    \EndProcedure
    \end{algorithmic}
    \end{algorithm}

    \item[]

    \item \textbf{LSTM-KM-RF}: This hybridization scheme advocates a classification perspective of the communication channel data processing pipeline, where LSTM feature outputs, optimized using K-means algorithm\cite{geng2019effective}, are fed to a Random Classifier regression \cite{sun2021hybrid}. The choice of Random Forest regression is widely motivated by its success in a lot of other alternative applications \cite{Breiman2001, djaballah2024hybrid,yan2024hybrid}. The architecture of this LSTM-KM-RF is detailed in \textbf{Algorithm \ref{alg:lstm_kmeans_rf}}.

    \begin{algorithm}[H]
    \caption{LSTM-KMeans-RandomForest Architecture}
    \label{alg:lstm_kmeans_rf}
    \begin{algorithmic}[1]
    \Procedure{Fit}{$X_{train}, y_{train}$}
        \State Train a primary LSTM model to predict $y_{train}$ from $X_{train}$.
        \State Generate LSTM predictions on the training data: $P_{lstm} \gets \text{LSTM.predict}(X_{train})$.
        \State Augment features with predictions: $X_{aug} \gets \text{concatenate}(X_{train}, P_{lstm})$.
        \State Train a K-Means model on $X_{aug}$ to generate cluster labels $C$.
        \State Further augment features with clusters: $X_{rf\_features} \gets \text{concatenate}(X_{aug}, C)$.
        \State C
        alculate LSTM residuals: $R_{train} \gets y_{train} - P_{lstm}$.
        \State Train a RandomForest model to predict $R_{train}$ from $X_{rf\_features}$.
    \EndProcedure
    \Procedure{Predict}{$X_{test}$}
        \State Generate primary predictions from the LSTM: $P_{lstm}$.
        \State Augment test features with predictions: $X_{aug} \gets \text{concatenate}(X_{test}, P_{lstm})$.
        \State Predict cluster labels using the trained K-Means model: $C \gets \text{KMeans.predict}(X_{aug})$.
        \State Construct final features for RF: $X_{rf\_features} \gets \text{concatenate}(X_{aug}, C)$.
        \State Predict residuals using the trained RandomForest: $P_{residuals} \gets \text{RF.predict}(X_{rf\_features})$.
        \State \textbf{return} $P_{lstm} + P_{residuals}$.
    \EndProcedure
    \end{algorithmic}
    \end{algorithm}

    \item[]

\end{itemize}

Each model processes disturbed signals and learns to reconstruct the ideal, undistorted communication trajectory. By operating in parallel and targeting the same supervised mapping, the three models (Prophet-LSTM, CNN-LSTM, LSTM-KM-RF)  serve as base learners within a stacked ensemble. A meta-learner, trained on the outputs of the base models, selects or combines their predictions to maximize accuracy. The training paradigm is designed to minimize the deviation between predicted and reference trajectories, thereby replicating the stabilizing function of wave variables while offering greater flexibility and performance in dynamic, uncertain environments. The ensemble model is constrained to operate within a passivity-preserving framework to ensure system stability.

\subsection{Stacking Ensemble Framework}
From the individual modules Prophet-LSTM, LSTM-KM-RF, CNN-LSTM, we can also question whether a potential combination of these modules would be beneficial. For this purpose, in this work, we propose a stacking ensemble model that uses a two-level learning strategy that combines the predictions of multiple diverse models to produce a final prediction that would be more accurate and generalizable than any single model. The architecture consists of Level-0 (base) learners and a Level-1 (meta) learner.

\subsubsection{Level-0: Base Learners}
The first level includes the three optimized hybrid models mentioned previously: the Prophet-LSTM, the LSTM-KMeans-RF, and the CNN-LSTM. Each of these models is trained independently on the full training dataset. Their role is to learn the mapping from both of the ideal and the disturbed input signals to the ideal, undistorted outputs from different "perspectives." More specifically, the Prophet-LSTM excels at capturing the underlying trends and seasonalities, while the CNN-LSTM is better at identifying localized spatio-temporal features. Finally, the LSTM-KMeans-Random Forest model leverages engineered features for a different regression approach.

\subsubsection{Level-1: Meta-Learner}
The second level consists of a single meta-learner. We chose the XGBoost Regressor \cite{XGBoost} for this role due to its computational efficiency and high performance with structured data, and its inherent robustness to overfitting, making it an ideal candidate to learn the complex mapping from base-learner predictions to the final output. Rather than training it on the original input data, the meta model is trained on the predictions of the Level-0 base learners instead, which are considered as "meta-features".

\begin{algorithm}[hb!]
    \caption{Meta-Feature Generation}
    \label{alg:meta-feature-generation}
    \begin{algorithmic}[1]
        \State \textbf{Input:} Training data $D_{train}$, Base learners $\{L_j\}_{j=1}^M$, Num. folds $K$.
        \State \textbf{Output:} Meta-features $X_{meta}$, Trained learners $\{L'_j\}_{j=1}^M$.
        \Statex
        \State Split $D_{train}$ into $K$ folds: $\{F_1, \dots, F_K\}$.
        \State Initialize meta-feature matrix $X_{meta}$.
        \For{each learner $L_j \in \{L_1, \dots, L_M\}$}
            \State \hspace{0.5cm} Initialize empty prediction vector $P_j$.
            \For{each fold $F_k \in \{F_1, \dots, F_K\}$}
                \State \hspace{1cm} $D_{k\_train} \leftarrow D_{train} \setminus F_k$.
                \State \hspace{1cm} $(X_{val\_k}, y_{val\_k}) \leftarrow F_k$.
                \State \hspace{1cm} Train: $L_{j,k} \leftarrow L_j(D_{k\_train})$.
                \State \hspace{1cm} Predict: $\hat{y}_{val\_k} \leftarrow L_{j,k}(X_{val\_k})$.
                \State \hspace{1cm} Store $\hat{y}_{val\_k}$ in $P_j$ at indices for $F_k$.
            \EndFor
            \State \hspace{0.5cm} Add $P_j$ as column(s) to $X_{meta}$.
            \State \hspace{0.5cm} Train on full data: $L'_j \leftarrow L_j(D_{train})$.
        \EndFor
        \State \textbf{return} $X_{meta}, \{L'_j\}_{j=1}^M$.
    \end{algorithmic}
\end{algorithm}

\begin{algorithm}[h!]
\caption{Stacking Ensemble Prediction}
\label{alg:stacking-prediction}
\begin{algorithmic}[1]
    \State \textbf{Input:} Test features $X_{test}$, Train labels $y_{train}$, Meta-features $X_{meta}$, Trained learners $\{L'_j\}_{j=1}^M$, Meta-learner $L_{meta}$.
\State \textbf{Output:} Final predictions $\hat{y}_{final}$.
    \Statex
    \State Train meta-learner: $L'_{meta} \leftarrow L_{meta}(X_{meta}, y_{train})$.
\Statex
    \State Generate test meta-features $X'_{meta}$:
    \State \hspace{0.5cm} Initialize empty matrix $X'_{meta}$.
\For{each learner $L'_j \in \{L'_1, \dots, L'_M\}$}
        \State Predict: $\hat{y}_{test\_j} \leftarrow L'_j(X_{test})$.
\State Add $\hat{y}_{test\_j}$ as column(s) to $X'_{meta}$.
    \EndFor
    \Statex
    \State Make final prediction: $\hat{y}_{final} \leftarrow L'_{meta}(X'_{meta})$.
\State \textbf{return} $\hat{y}_{final}$.
\end{algorithmic}
\end{algorithm}

The process, as outlined in \textbf{Algorithm} \textbf{\autoref{alg:meta-feature-generation} }and \textbf{Algorithm} \textbf{\autoref{alg:stacking-prediction}}, works as such:

\begin{enumerate}
    \item Meta-Feature Generation: In order to prevent information leakage and  generate a robust training set for the meta-learner, we use K-fold cross-validation. The training data (85\% of the original data) is split into K folds. For each fold, the three base models are trained on the remaining K-1 folds, and then make predictions on the held-out fold. These "out-of-fold" predictions are collected and form the feature set (meta-features) that will be later used to train the meta-learner.

    \item Meta-Learner Training: Once the meta-features are generated for the entire training set, the meta-learner (XGBoost) is trained to map these meta-features to the real output values. In other words, the meta-learner learns the optimal way to combine the predictions of the base models.

    \item Final Prediction: To generate predictions on the unseen test data (15\% of the original dataset), the inputs are initially passed through all three base models, which have been previously retrained on the full training dataset. Their predictions are then fed as an input to the trained meta-learner, which will generate the final output predictions.
\end{enumerate}

This stacking approach allows the system to leverage the strengths of each base model, resulting in a more powerful and resilient predictive unit to replace the traditional wave-variable transform.

\subsection{Model Optimization and Performance Tuning}
To ensure a fair comparison and maximize performance, all base models were optimized using Optuna \cite{Akiba2019}, a modern hyperparameter optimization library based on Tree-structured Parzen Estimators. Each model's search space included key architectural and training parameters (e.g., number of LSTM units, learning rate, dropout, batch size). We used validation RMSE as an objective functions to guide the optimization. By including ensemble learning and systematic hyperparameter tuning, the proposed stacked hybrid architecture matches the performance of traditional wave-variable-based communication links, providing a promising solution for teleoperation systems operating under non-ideal, variable-delay conditions.

\subsection{Performance Evaluation Metrics and Procedure}
A comprehensive methodology was adopted to rigorously evaluate the four models. The evaluation encompassed quantitative accuracy metrics, passivity and stability assessments, and qualitative visual inspections to ensure the models' viability in the bilateral teleoperation context.
\subsubsection{Quantitative Metrics}
Each model was assessed using standard regression metrics to evaluate predictive accuracy:
\begin{itemize}
    \item \textbf{Root Mean Squared Error (RMSE)} – captures the typical magnitude of the prediction errors.
\item \textbf{Mean Absolute Error (MAE)} and \textbf{Mean Squared Error (MSE)} – measure the average prediction errors.
\item \textbf{R-squared (R²)} – quantifies the proportion of variance in the output explained by the model.
\item \textbf{Training and Tuning Times} – assess the computational efficiency of each configuration.
\end{itemize}

\subsubsection{Passivity Validation}
Maintaining the passivity of the communication channel is essential for system stability in bilateral teleoperation. Passivity describes the energy behavior of a system: system is considered passive if it doesn't generate its own energy, it only stores or dissipates the energy supplied to it \cite{khalil}. In the context of machine learning, especially for models used in dynamic or control applications, passivity is also a strong guarantee of stability. Verifying this property ensures that the model will behave predictably when deployed, without producing runaway outputs\cite{khalil}. The condition for passivity for a model $f$ having an input $u$ and output $y=f(u)$ is:
\begin{equation}
    \int_0^T u(t)^T y(t) \, dt \geq 0 \quad \text{(for zero initial energy)}
    \label{eq:integral_inequality}
\end{equation}

\subsubsection{Stability Validation}

To complement passivity analysis, the Lipschitz constant of the model is estimated to evaluate its intrinsic stability. The Lipschitz constant quantifies the maximum amplification factor of the network, providing a bound on how much the output can change relative to a change in the input. A smaller Lipschitz constant indicates that the network is less prone to amplifying disturbances, which is crucial to maintain the stability of the overall system. For a function $f:\mathcal{X} \rightarrow \mathcal{Y}$, the Lipschitz constant $L$ is defined as:
\begin{equation}
    \|f(x_1) - f(x_2)\| \le L \|x_1 - x_2\| \quad \forall x_1, x_2 \in \mathcal{X}
\end{equation}
where $\|\cdot\|$ denotes a chosen norm (e.g., Euclidean norm).

In neural networks, $L$ corresponds to the largest singular value of the Jacobian matrix, indicating the maximum gain. A smaller $L$ implies reduced sensitivity to input perturbations.

\begin{algorithm}[b!]
\caption{Meta-Model Stability Analysis}
\label{alg:stability}
\begin{algorithmic}[1]
\State \textbf{Input:} Trained meta-model $M_{meta}$, test meta-features $X_{meta\_test}$
\State \textbf{Output:} Lipschitz constant estimate $L$, Passivity ratio $\rho_{passive}$

\Procedure{AnalyzeStability}{$M_{meta}, X_{meta\_test}$}
    \State {\textit{\textbf{Part 1: Estimate Lipschitz Constant}}}
\State Initialize $L_{max} \gets 0$.
\For{a number of iterations}
    \State Randomly sample two distinct points $x_1, x_2$ from $X_{meta\_test}$.
\State Calculate input distance: $d_{in} = \|x_1 - x_2\|_2$.
    \State Generate predictions: $p_1 \gets M_{meta}.\text{predict}(x_1)$, $p_2 \gets M_{meta}.\text{predict}(x_2)$.
\State Calculate output distance: $d_{out} = \|p_1 - p_2\|_2$.
    \State If $d_{in} > 0$, update $L_{max} \gets \max(L_{max}, d_{out} / d_{in})$.
\EndFor
\State $L \gets L_{max}$.
\State Provide an observation based on the magnitude of $L$.
\State {\textit{\textbf{Part 2: Verify Passivity (Non-Expansiveness)}}}
\State Generate predictions for all test data: $Y_{pred} \gets M_{meta}.\text{predict}(X_{meta\_test})$.
\State Calculate norms for each sample: $N_{in} = \|X_{meta\_test}\|_2$, $N_{out} = \|Y_{pred}\|_2$.
\State Count non-expansive samples: $C_{passive} \gets \sum_{i} [N_{out,i} \le N_{in,i}]$.
\State Calculate passivity ratio: $\rho_{passive} \gets (C_{passive} / \text{len}(X_{meta\_test})) \times 100$.
\State Provide an observation based on the value of $\rho_{passive}$.
\State \textbf{return} $L, \rho_{passive}$
\EndProcedure
\end{algorithmic}
\end{algorithm}

\begin{table*}[t!]
\centering
\small 
\setlength{\tabcolsep}{4pt} 
\renewcommand{\arraystretch}{0.9} 
\caption{Overview of the Stacking Ensemble Model Architectures and Hyperparameters.}
\label{tab:model_setup}
\begin{tabular*}{\textwidth}{@{\extracolsep{\fill}} 
    >{\RaggedRight\arraybackslash}p{2.5cm}
    >{\RaggedRight\arraybackslash}p{4cm}
    >{\RaggedRight\arraybackslash}p{6.5cm}
    >{\RaggedRight\arraybackslash}p{2.5cm}
@{} }
\toprule
\textbf{Abbr.} & \textbf{Model Component} & \textbf{Key Hyperparameters} & \textbf{Primary Libraries} \\
\midrule

\multirow{2}{*}{\centering Prophet-LSTM} & Prophet (trend \& seasonality) & \texttt{seasonality\_mode='multiplicative'} & \multirow{2}{2.5cm}{\RaggedRight Prophet, TensorFlow} \\
& LSTM (for residuals) & \texttt{units=64, dropout=0.3, activation='tanh'} & \\
\midrule

\multirow{3}{*}{\centering LSTM-KMeans-RF} & LSTM (initial prediction) & \texttt{units=64, dropout=0.3, sequence\_length=100} &
\multirow{3}{2.5cm}{\RaggedRight TensorFlow, Scikit-learn\cite{scikit}} \\
& K-Means (clustering) & \texttt{n\_clusters=5} & \\
& Random Forest (residuals) & \texttt{n\_estimators=150, max\_depth=20} & \\
\midrule

\multirow{2}{*}{\centering CNN-LSTM} & 1D-CNN (feature extraction) & \texttt{filters=64, kernel\_size=5} & \multirow{2}{2.5cm}{\RaggedRight TensorFlow} \\
& LSTM (sequence modeling) & \texttt{units=100, dropout=0.3} & \\
\midrule

\multirow{2}{*}{\centering Meta-XGB} & XGBoost Regressor & \texttt{n\_estimators=200, learning\_rate=0.05} & \multirow{2}{2.5cm}{\RaggedRight XGBoost, Scikit-learn\cite{scikit}} \\
& (as meta-learner) & \texttt{max\_depth=5, subsample=0.8} & \\

\bottomrule
\end{tabular*}
\end{table*}

For a learned model based communication:
\begin{itemize}
    \item \textbf{Stability}: $L \le 1$ limits the amplification of disturbances, preventing thereby oscillatory or unstable behavior in the closed-loop teleoperation system.
    \item \textbf{Robustness}: The Lipschitz property is the mathematical foundation for robustness against input perturbations. Formally, it guarantees that for any input $x$  and a perturbed version $x+\delta$, the output deviation is bounded:  \begin{equation}
        \| f(x) - f(x + \delta) \| \leq L \cdot \| \delta \|
        \label{eq:placeholder}
    \end{equation}

    A small value of L directly implies that minor input variations—such as sensor noise, communication jitter, or slight deviations in operator input—will only lead to proportionally small changes in the network's output. This prevents the model from reacting erratically to insignificant fluctuations, thereby enhancing its reliability and ensuring the predictability needed for safe physical interaction.
    
    \item \textbf{Passivity Proxy}: Although not a formal passivity metric, $L \approx 1$ presents a strong passivity proxy. It suggests that the network does not generate energy, as the output's magnitude is bounded by the input's magnitude. 
    
    This behavior is analogous to that of a passive physical system, which is a desirable characteristic for stable haptic feedback.
\end{itemize}

The power iteration method \cite{pow} is used to estimate $L$, offering a tractable approach for analyzing complex, non-linear learned models. The \textbf{Algorithm \textbf{\autoref{alg:stability} }}details both of the stability and passivity evaluations of our meta model.

\subsubsection{Qualitative Comparison (Transparency Validation)}

A visual analysis was performed by overlaying our meta model's predictions, as well as each of the level 0 hybrid models, with the targeted ideal responses of the original baseline system. These time-domain comparisons assessed the extent to which each model preserved transparency, responsiveness and robustness in the presence of variable random time delays, and additional white noise as an additive perturbation at the input ports of the communication channels.

This qualitative inspection complements the numerical analysis by revealing subtle performance differences not easily captured by scalar metrics.

\section{Experimental Setup}

The simulation environment was implemented in MATLAB/Simulink. Both the master and slave devices are modeled as identical single-degree-of-freedom (1-DoF) robots, represented by their respective impedance models:

\begin{equation}
    Z_m = Z_s = 0.25s + 0.8
\end{equation}

These impedances relate velocity to force $(V/F)$ at each terminal. The local controllers $C_m$ and $C_s$ are implemented to ensure the position stability of the master and slave devices, respectively. Their transfer functions are defined as:

\begin{equation}
    C_m = C_s = 0.0449 + \frac{1}{s}
\end{equation}

Following the optimization procedure outlined in \cite{mitiche23}, the parameters of the controllers for the four-channel teleoperation architecture were selected to achieve transparency and stability in the presence of communication delays.

The resulting architecture parameters are as follows:

\begin{align}
    C_1 &= -C_4 = 1 + \frac{1}{s}, \quad C_2 = 0.5, \quad C_3 = 2.1790 \times 10^{-5} \\
    C_5 &= -5.8165, \quad C_6 = 0.0038, \quad b = 62.1208
\end{align}

This configuration ensures that the nominal system operates stably and passively without the data-driven LSTM component. 

The simulation data was generated by subjecting the baseline system to a variety of input force signals $F^*_h$ from the human operator, including step, sinusoidal, and complex spline signals, to capture a wide range of operational dynamics. To create a robust dataset and train a disturbance-resilient model, each unique input signal was used to generate two distinct response scenarios. First, an ideal scenario was run with no additional noise with the system's 2-port communication channel responses ($M_1$\textbf{[N]}, $M_2$\textbf{[N]}, $N_1$\textbf{[N]}, $N_2$\textbf{[N]}) recorded as the ground truth. Secondly, a disturbed scenario was run using the exact same input signals but introducing significant real-world challenges: variable time delays $T_d$, simulated using the \textit{random} block in Simulink with a Gaussian distribution (with a mean value of 0, a variance of 0.001 and a sample time of 2 seconds), and additive white noise (simulated with a \textit{Band-Limited White Noise} block in Simulink, with a noise power of 1 and a sample time of 0.1s).  injected into the input channels ($M_1$ and $N_2$) to mimic unpredictable signal degradation.

\begin{figure}[t!]
    \centering
    \includegraphics[width=\linewidth]{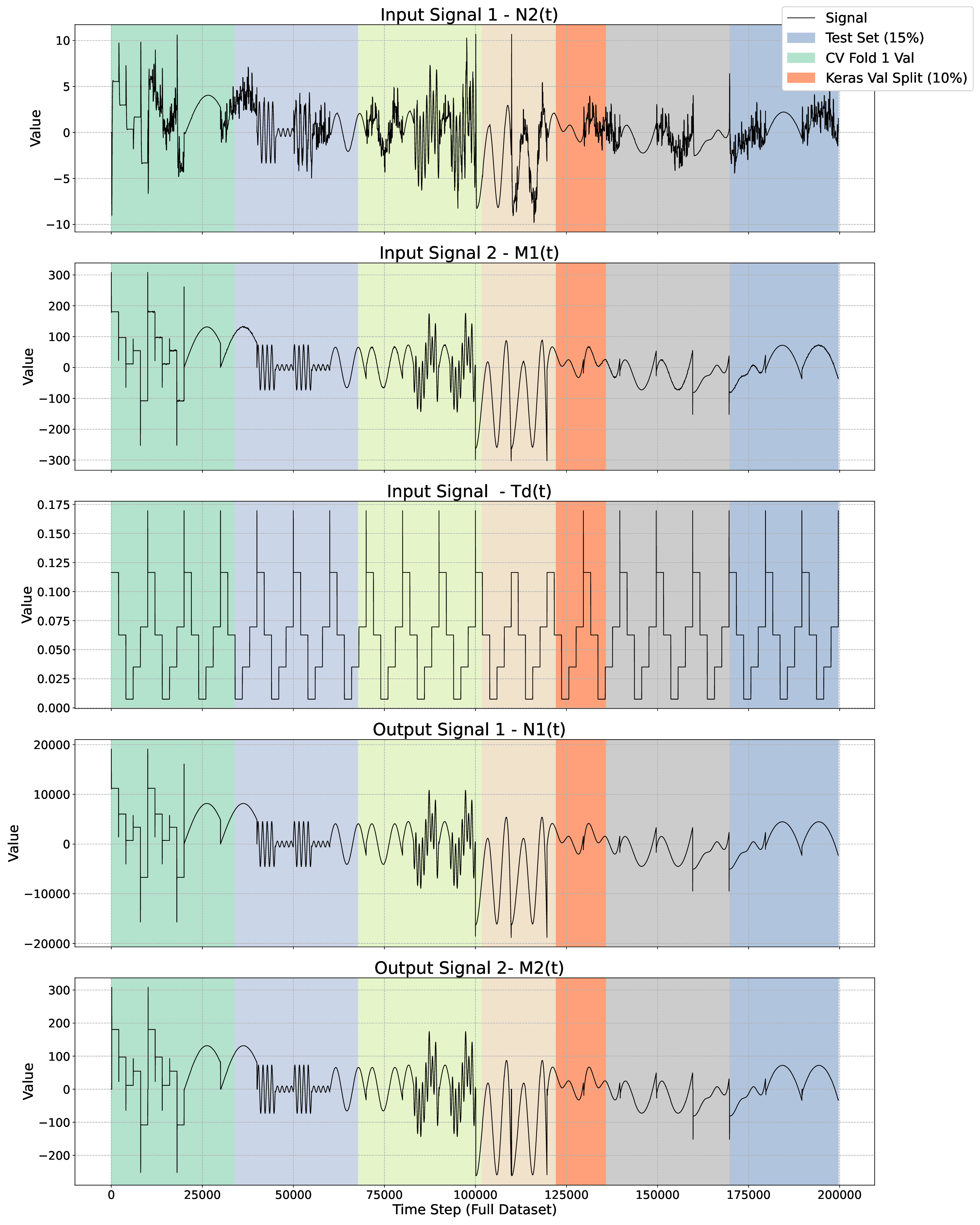}
    \caption{Visualization of the dataset splits across the full time-domain signals for all inputs and outputs. The main division into the Training Set (85\%) and Test Set (15\%, light blue) is shown. Within the training set, the 5-fold cross-validation splits are depicted by pastel-colored vertical bands. Finally, the internal 10\% validation set used for training each neural network (e.g., for early stopping) is shown with a hatched red overlay.}
    \label{fig:full_data}
\end{figure}

As illustrated in\textbf{ \autoref{fig:full_data}}, the dataset comprises these paired signals. The inputs for our machine learning framework are the disturbed channel signals ($M_1$, $N_2$) and the variable time delay profile $T_d$.  The target outputs for supervised learning are the corresponding ideal, non-noisy responses ($N_1$, $M_2$) from the baseline system. This approach explicitly trains the ensemble model to learn a mapping from a noisy, unpredictably delayed input to a clean, stable output, effectively teaching it to function as a robust filter and delay compensator. The complete dataset was then preprocessed—normalized, cleaned of outliers, and windowed into sequences of 100 timesteps, and finally split into a training set (85\%) and a test set (15\%) , with the training set further divided using 5-fold cross-validation for robust meta-feature generation.

\vspace{2pt}

The optimal set of the base models hyperparameters, as well as those of the meta model, are detailed in the \textbf{\autoref{tab:model_setup}}.

\section{Results and Discussion}
This section details the quantitative and qualitative performance of the proposed data-driven ensemble. We evaluate the predictive accuracy of the base learners and the final meta-model, analyze their computational cost, assess the system's stability, and interpret the overall findings.

\subsection{Performance Metrics}
The predictive performance of each base model and the final XGBoost meta-model was evaluated on the test set using RMSE, MAE, and R² metrics, as detailed in \textbf{\autoref{tab:Table2}}. 

\begin{table}[h]
    \centering
    \scriptsize 
    \setlength{\tabcolsep}{2pt} 
    \renewcommand{\arraystretch}{0.9} 
    \caption{Performance Comparison}
    \label{tab:Table2}
    \begin{tabularx}{0.95\linewidth}{ 
        >{\centering\arraybackslash}X 
        *{6}{c} 
    }
        \toprule
        
        \multirow{2}{*}{\textbf{Method}} & \multicolumn{2}{c}{\textbf{RMSE}} & \multicolumn{2}{c}{\textbf{MAE}} & \multicolumn{2}{c}{\textbf{R\(^2\)}} \\
        
        \cmidrule(lr){2-3} \cmidrule(lr){4-5} \cmidrule(lr){6-7}
        
        & \textbf{N1} & \textbf{M2} & \textbf{N1} & \textbf{M2} & \textbf{N1} & \textbf{M2} \\
        \midrule
        LSTM+Prophet & 1.372\% & 15.798\% & 0.922\% & 8.870\% & 99.984\% & 97.929\% \\
        \midrule
        
        LSTM+K-means+RF & 1.725\% & 11.537\% & 0.973\% & 5.068\% & 99.975\% & 98.896\% \\
        \midrule
        LSTM+CNN & 9.041\% & 13.198\% & 7.833\% & 5.969\% & 99.309\% & 98.554\% \\
        \midrule
        Meta Model (XGBoost) & 1.994\% & 10.635\% & 1.320\% & 4.988\% & 99.966\% & 99.062\% \\
        \bottomrule
    \end{tabularx}
\end{table}

Among the base models, the Prophet-LSTM model achieved the best metrics for Output 1 ($N_1$) with a RMSE of 1.372\%, a MAE of 0.922\% and a $R^2$ of 99.984\%, whereas for Output 2 ($M_2$), the LSTM-Kmeans-RF model achieved the best metrics with a RMSE of 11.537\% , a MAE of 5.068\% and a $R^2$ of 98.896\%. This suggests that different architectures may be more effective in capturing the dynamics of distinct signal components, justifying the ensemble approach.

The final Meta-Model (XGBoost) achieves a performance that is highly competitive with the best individual models, achieving R² values of 99.966\% and 99.062\% for the two outputs, respectively. While it does not outperform every single best model on every metric for both outputs, it provides consistently enhanced results for the Output 2 ($M_2$), indicating a robust, generalized performance by effectively blending the strengths of the base learners. This confirms the effectiveness of the stacking strategy in creating a balanced and highly accurate predictive model.

\textbf{\autoref{tab:Table3}} details the training and inference durations. The CNN-LSTM was the fastest to train (102.2 sec), while the LSTM+K-means+RF was the most computationally intensive (1332.0 sec). The total training time of approximately 44 minutes represents a one-time, offline cost. The meta-model itself adds negligible overhead, with a training time of only 4.7 sec and near-instantaneous inference time (0.006 sec), making it very suitable for real-time application.

\begin{table}[h!]
    \centering
    \scriptsize 
    \setlength{\tabcolsep}{4pt} 
    \renewcommand{\arraystretch}{0.9} 
    \caption{Training Durations}
    \label{tab:Table3}
    \begin{tabularx}{0.8\linewidth}{ 
        >{\raggedright\arraybackslash}X 
        c  
        c  
    }
        \toprule
        \textbf{Method} & \textbf{Train Time (sec)} & \textbf{Inference Time (sec)} \\
        \midrule
        LSTM+Prophet & 1172.989 & 11.362 \\
        \midrule
        LSTM+K-means+RF & 1332.018 & 3.004 \\
        \midrule
        LSTM+CNN & 102.206 & 1.934 \\
        \midrule
        Meta Model (XGBoost) & 4.684 & 0.006 \\
        \midrule
        \textbf{Total Training Time} & \multicolumn{2}{c}{\textbf{2611.899}} \\
        \bottomrule
    \end{tabularx}
\end{table}

\subsection{Stability and Passivity Analysis}
The passivity validation, shown in the \textbf{\autoref{tab:stability_analysis_col}}, as a sufficient but not necessary condition, yielded a passivity ratio of 100\%, which means that for the totality of the test data, the model behaves as a passive system, dissipating or storing energy rather than generating it. This adherence to passivity is crucial for preventing oscillations and guaranteeing the stability of the overall teleoperation system. For a more robust stability evaluation, we estimated the Lipschitz constant for the final meta-model, which was found to be equal to 0.4467, which is less than 1: it means that the model will not amplify disturbances, thereby ensuring the intrinsic stability of the communication channel.

\begin{table}[H]

\small
\centering
\caption{Stability and Passivity Analysis of the Meta-Model.}
\label{tab:stability_analysis_col}

\begin{tabularx}{\columnwidth}{@{} l c X @{}}
\toprule
\textbf{Metric} & \textbf{Value} & \textbf{Observation} \\
\midrule

\multicolumn{3}{l}{\textit{Lipschitz Analysis}} \\
\hspace{5pt}Avg. Lipschitz ($L$) & 0.4467 & Contractive ($L<1$); ensures stability by preventing disturbance amplification. \\

\addlinespace[2pt]
\hspace{5pt}$L$ for Output 1 & 0.860 &  Shows higher sensitivity to input perturbations, but remains stable. \\
\addlinespace[2pt]
\hspace{5pt}$L$ for Output 2 & 0.033 & Highly stable behavior for this channel. \\
\midrule
\multicolumn{3}{l}{\textit{Passivity Analysis}} \\
\hspace{5pt}Metric Used & $||.||_{sample}$ & Non-Expansiveness ($||y|| \le ||u||$) used as a sufficient condition. \\
\addlinespace[2pt]
\hspace{5pt}Passivity Ratio & 100\% & Strongly non-expansive; guarantees passive behavior for all 29948 test samples. \\
\bottomrule
\end{tabularx}
\end{table}

\subsection{Interpretations and Visual Analysis}

The quantitative results in and the stability analysis in \autoref{tab:stability_analysis_col} provide a multi-faceted view of the proposed framework's performance. A deeper analysis reveals important insights into the behavior of the individual models and the effectiveness of the stacking strategy. 

\subsubsection{Specialization of Base Learners}
The quantitative results in \autoref{tab:Table2}  provide a clear justification for our ensemble strategy. We observe a distinct model specialization: the Prophet-LSTM model excelled on Output 1 (RMSE: 1.372\%), while the LSTM-KMeans-RF model performed best on the more challenging Output 2 (RMSE: 11.537\%).

\begin{figure}[t!]
    \centering
    \includegraphics[width=\linewidth]{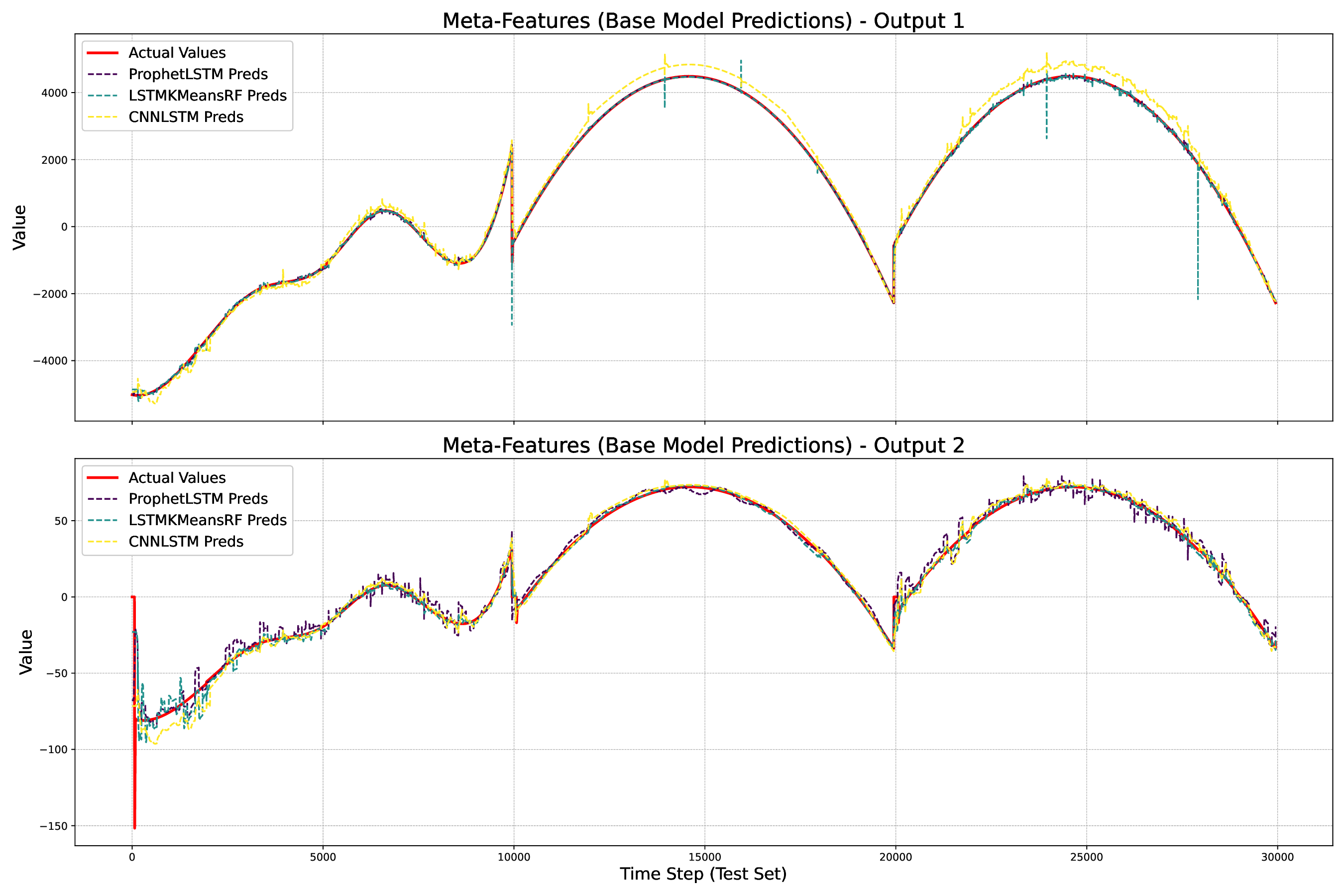}
    \caption{Predictions from base learners (Prophet-LSTM, LSTM-KMeans-RF, and CNN-LSTM) compared to ground truth for Output 1 (N1) and Output 2 (M2). These serve as meta-features for the final ensemble model.}
    \label{fig:base_plots}
\end{figure}

This performance disparity can be understood by visually inspecting the signals. As seen in the plots of \textbf{\autoref{fig:base_plots}}, Output 1 is characterized by smoother, more predictable trend-based dynamics. The Prophet-LSTM model, designed specifically to decompose and model such trends, was naturally suited for this task . 

In contrast, Output 2 exhibits significantly more noise and high-frequency volatility. Here, the feature-based approach of the LSTM-KMeans-RF model proved superior. Its ability to identify distinct operating regimes with K-Means and apply a robust Random Forest regressor allowed it to generalize more effectively in the presence of noise that could confound a pure sequence-to-sequence model .

\subsubsection{Stacking Ensemble Effect}
This observed specialization highlights the limitations of a one-size-fits-all approach. The true strength of our framework is revealed by the Meta-Model's performance. It is not designed to simply be the best on every metric, but to provide the best generalized performance by intelligently blending the outputs of the specialist models . 

\begin{figure}[h!]
    \centering
    \includegraphics[width=\linewidth]{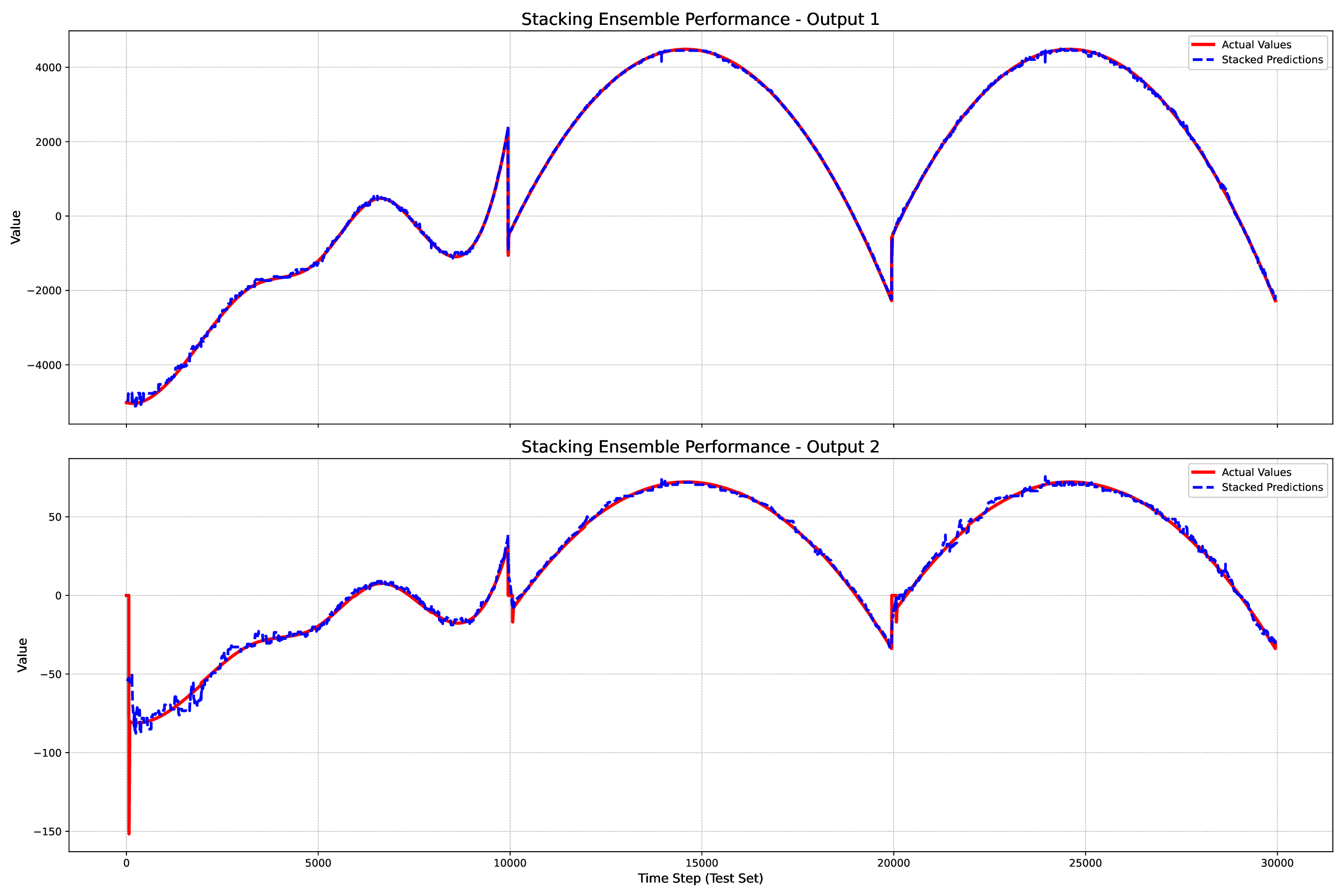}
    \caption{Meta-Learner's Final Predictions on the Test Dataset}
    \label{fig:stacking_plots}
\end{figure}

This is confirmed by the results for Output 2, where the meta-model's predictions were better than those of the best individual base models on every single metric, all while giving competitively good results for Output 1 as well. This demonstrates that the meta-learner is not merely averaging predictions, but is actively leveraging the strengths of each base learner to create a more robust and accurate final output, indicating a successful stacked ensemble, as we can observe in \textbf{\autoref{fig:stacking_plots}}.

\begin{figure}[h]
    \centering
    \includegraphics[width=\linewidth]{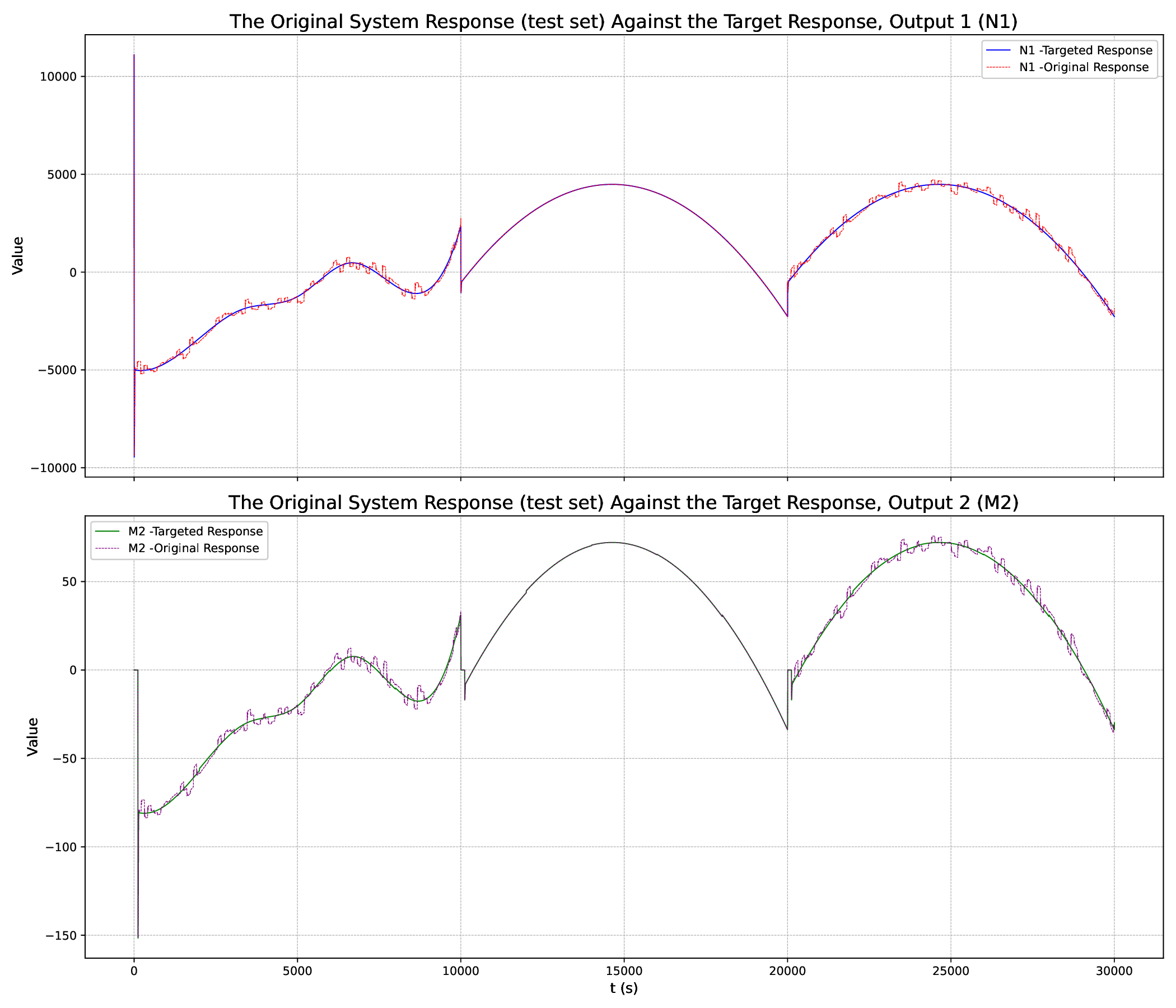}
    \caption{Visualization of the original system's response over the test dataset (15\%) using Chen's architecture, compared to the targeted responses fed to the learning framework.}
    \label{fig:chen_plots}
\end{figure}

Our model's predictions also show a noticeable improvement compared to the responses of the baseline wave-variable based model in the same delay and noises conditions, as shown in\textbf{ \autoref{fig:chen_plots}}, confirming again our stacked ensemble performance.

\subsubsection{Stability as a Core, Verifiable Feature}
Beyond accuracy, the stability and passivity analysis detailed in \autoref{tab:Table3} confirms the viability of this data-driven model in a control context. An estimated average Lipschitz constant of 0.4467 (well below the stability boundary of 1) and a 100\% passivity ratio on the test data are powerful guarantees. It demonstrates that a complex, non-linear model can be designed and constrained to behave predictably, preventing the amplification of disturbances—a non-negotiable requirement for physical human-robot interaction. The fact that Output 1's Lipschitz constant (0.860) is higher than Output 2's (0.033) aligns with our qualitative visual analysis, indicating that this channel is inherently more sensitive, yet the model successfully keeps it within stable bounds.

\section{Conclusion and future works}
This article addressed the challenge of maintaining stability and transparency in bilateral teleoperation systems subject to time delays and communication disturbances. We proposed a novel data-driven framework that replaces the conventional wave-variable transform in a four-channel architecture with a robust stacking ensemble of hybrid deep learning models. The ensemble, composed of a Prophet-LSTM, an LSTM-KMeans-RF, and a CNN-LSTM as base learners, with an XGBoost meta-learner, was designed to learn the ideal communication dynamics from data corrupted by variable delays and noise.

Our experimental results, conducted on a high-fidelity MATLAB/Simulink simulation, demonstrate that the proposed ensemble achieves excellent predictive accuracy, with R² values exceeding 99\%. More importantly, we formally verified the stability of the learned model. With an estimated Lipschitz constant of 0.4467 and a passivity ratio of 100\% of the samples, our framework guarantees that it does not amplify disturbances and adheres to the passivity constraints essential for safe teleoperation. The final model proved capable of matching the transparency of the baseline wave-variable system while offering enhanced robustness against noise.

This work successfully demonstrates that a carefully designed and constrained machine learning model can serve as a viable, high-performing alternative to traditional control components in stability-critical applications. This opens up several potentials in robotics and man-machine systems both in civilian and military fields \cite{Adib2023}.

For future work, we identify three main directions. First, the current framework should be validated on a physical, hardware-in-the-loop experimental setup to assess its performance in a real-world environment. Second, the approach could be extended to multi-degree-of-freedom (multi-DoF) teleoperation systems, which present more complex dynamics. Finally, exploring online adaptation mechanisms for the meta-learner could further enhance the system's ability to cope with dynamically changing network conditions and environments.

\bibliographystyle{elsarticle-num}
\bibliography{cleaned}

\end{document}